# Comparison of the Information Technology Development in Slovakia and Hungary

Peter Sasvari  
Institute of Business Sciences, Faculty of Economics  
University of Miskolc  
Miskolc-Egyetemvaros, Hungary

Zsuzsa Majoros  
Institute of Business Sciences, Faculty of Economics  
University of Miskolc  
Miskolc-Egyetemvaros, Hungary

*Abstract*— Nowadays the role of information is increasingly important, so every company has to provide the efficient procurement, processing, storage and visualization of this special resource in hope to stay competitive. More and more enterprises introduce Enterprise Resource Planning System to be able to perform the listed functions. The article illustrates the usage of these systems in Hungary and Slovakia, as well as tests the following presumption: the level of Information Technology (IT) development is lower in Hungary than our northern neighbor.

*Keywords*— *Information society; Information Technology; Slovakia, Hungary*[1]

## I. INTRODUCTION

The role of information has become more and more substantial in the economy recently, and information is regarded as an important resource since it is more difficult for companies to improve their market positions in the long term without having the appropriate amount of available information [5]. Globalization in the business world has brought about the possibility of getting a greater amount of information in much less time, which means that companies are forced to spend more time and energy on handling the increased information load [10] [17].

As information systems are designed to provide effective help in this process, they are becoming increasingly popular among companies due to the robust technological development [11]. This paper deals with the usage of business information systems among Hungarian enterprises and analyzes the following three key questions: how the usage of business information systems influences a company's economic performance, what expenditure is required for an individual company to develop its information technology infrastructure and finally, to what extent information technology is considered important as a functional area within the organization of a company [2].

The aim of the research presented in this paper was to explore the current situation of Hungarian enterprises in terms of using business information systems, gaining a more thorough insight into the background of the decisions made on introducing such information systems, together with the possible problems related to their introduction and further usage [10].

The IT development of the two countries compared (Hungary (HUN) and Slovakia (SK)) may be corresponding with their economic status, so first we have analyzed some essential economic indicators.

TABLE I.  ECONOMIC INDICATORS IN HUNGARY AND SLOVAKIA [27] [28]

|  | 2008 | | 2009 | | 2010 | | 2011 | |
|---|---|---|---|---|---|---|---|---|
|  | HUN | SK | HUN | SK | HUN | SK | HUN | SK |
| GDP (PPP) EU=100 (%) | 64 | 73 | 65 | 73 | 65 | 73 | 66 | 73 |
| Unemployment rate (%) | 7.8 | 9.6 | 10.0 | 12.1 | 11.2 | 14.4 | 10.9 | 13.5 |
| Avarage monthly gross wage (EUR) | 705.5 | 723.0 | 708.5 | 744.5 | 718.4 | 769.0 | 755.6 | 786.0 |

The GDP /purchasing power parity/ indicate the quality of living standards; this data is constantly higher in Slovakia, while in Hungary we can see a slow development. The unemployment rate is lower in our country, so this indicator favors us. However the average monthly gross wage is greater in Slovakia.

In recent years the number of companies has multiplied across the EU, only the economic crisis has sat back this dynamism. Micro enterprises make up the most typical size both in Europe and in Hungary, this amount to 91.8% of all the companies. Since most jobs are created by SME-s, we cannot ignore researching them. In Hungary the number of SME-s per 1000 habitants is higher than the average in the EU. The reason of this is the existence of forced businesses, formed by the regulatory environment.

Hungary was the 49[th] in 2012 by the classification of the Doing Business Index, while Slovakia was the 46[th]. This indicator scans different parameters in the economies then shows their ease of doing business. In 2013 Slovakia keeps its position, but Hungary becomes only the 54[th], because of the difficulty in access to credit, the deterioration in solving bankruptcies, and the hardness of starting business.

[1] The described work was carried out as part of the TÁMOP-4.2.1.B-10/2/KONV-2010-0001 project in the framework of the New Hungarian Development Plan. The realization of this project is supported by the European Union, co-financed by the European Social Fund.

Whereas these economic indicators show that Slovakia is in a more favorable economic position. Our purpose is to analyze that question of „which country is more developed by IT perspective".

## II. Definition and classification of Information Systems

There are several definitions offered on business information systems in the literature. According to Burt and Taylor's approach, "business information systems can be regarded as an information source in any combination thereof, or any access to and any recovery of their use or manipulation. Any business information system is designed to link the user to an appropriate source of information that the user actually needs, with the expectation that the user will be able to access the information satisfying their needs" [3]. Davis and Olson define business information systems as "an integrated user-machine system for providing information to support the operations, management, analysis, and decision-making functions in an organization. The system utilizes computer hardware and software, manual procedures, models for analysis, planning, control, and decision-making by using a database" [6].

"Information systems are a part of any organization that provides, generates, stores, separates, divides and uses information. They are made up of human, technical, financial and economic components and resources. In fact, they can be regarded as inherently human systems (organizations, manual systems) that may include a computer system, and automatizes certain well-defined parts and selected items of the system. Its aim is to support both the management functions and the daily operation of an organization." [7]

In a broader sense, a business information system is the collection of individuals, activities and equipment employed to collect, process and store information related to the company's environment, its internal activities, together with all transactions between the company and its environment. Beyond giving direct support to operations, its basic task is to provide decision-makers with the necessary information during the whole decision-making process. The system's main components are the following [9]:

- Individuals carrying out corporate activities: the actual users of technical apparatus. Decision-makers also belong to this group, as leaders who receive information on the factors affecting business operations, and use business information systems to make decisions in relation to planning, implementation and monitoring business activities.
- Information (also known as processed data on external and internal facts) which – due to its systematized form – can be used directly in the decision-making process.
- Technical apparatus, nowadays usually a computer system that supports and connects the subsystems applied to achieve corporate objectives.

The computer system standardizes a significant part of the information and communication system, thus making it easier to produce and use information.

According to one definition proposed [4] "information systems are systems that use information technology to collect information, transmit, store, retrieve, process, display and transform information in a business organization by using information technology."

Raffai's understanding of information systems is as follows: "it uses data and information as a basic resource for different processing activities in order to provide useful information for performing useful organizational tasks. It's main purpose is the production of information, that is dedicated to creating messages that are new to the user, uncertainties persist, and their duties, to assist in fulfilling the decisions" [19].

The classification of business information systems is a difficult task because, due to the continuous development, it is hard to find a classification system that can present unanimously defined information system types. It occurs quite often that different abbreviations are used to refer to the same system or certain system types appear to be merged together. As a consequence, the classification of business information systems can be done in several ways, the lists of several groups of business information systems presented below just to show a few alternatives for classification [1].

Dobay [8] made a distinction between the following types:

- Office Automation Systems (OAS): used for efficient handling of personal and organizational data (text, image, number, voice), making calculations and document management.
- Communication systems: supporting the information flow between groups of people in a wide variety of forms.
- Transaction-processing systems (TPS): used for receiving the initiated signals of transactions, generating and giving feedback on the transaction event.
- Management Information Systems (MIS): used for transforming TPS-related data into information for controlling, management and analysis purposes.
- Executive Information Systems (EIS): intended to give well-structured, aggregated information for decision-making purposes.
- Decision support systems (DSS): applied to support decision-making processes with information, modelling tools and analytical methods.
- Facility Management Systems (facility management, production management): used for directly supporting the value production process.
- Group work systems: intended to give group access to data files, to facilitate structured workflows and the implementation of work schedules.

Another possible approach to defining categories is based on Raffai's work [19]:

- Implementation support systems: this group includes transaction processing systems (TPS), process control

(IJACSA) International Journal of Advanced Computer Science and Applications,
Vol. 4, No. 2, 2013

systems (PCS), online transaction processing systems (OLTP), office automation systems (OAS), group work support systems (GS), workflow management (WF), and customer relation management systems (CRM).

- Executive work support systems: this category can include strategic information systems (SIS), executive information systems (EIS), online analytical processing systems (OLAP), decision support systems (DSS), group decision support systems (GDSS), and management information systems (MIS).

- Other support systems: business support systems, (BIS), expert systems (ES), integrated information processing systems (IIS), and inter-organizational information systems (IOS) can be found in this category.

Based on Gábor's [12] findings, information systems can also be examined by applying the following classification criteria.

- According to organizational structure:
    o functional systems such as reporting applications,
    o comprehensive business systems such as corporate management systems used by the entire organization,
    o inter-organizational systems such as reservation systems.

- According to the field of application:

Depending on the scope of activities, systems used for accounting, finance, production, marketing or human resource management belong to this category. These systems are generally related to the various functions a company performs.

- According to the type of support:
    o TPS (Transaction Processing System) – it focuses on a particular purpose, its basic function is to serve as a supporting tool for data processing related to business activities.
    o MIS (Management Information System) – it basically supports functional executive activities (O'Brien 1999).
    o KMS (Knowledge Management System) – it facilitates the execution of tasks related to knowledge as a valuable corporate resource.
    o OAS (Office Automation System) – it supports office document management, group work and communication.
    o DSS (Decision Support System) – it supports decisions made by managers and analyses done by experts.
    o EIS (Enterprise Information System) – it is designed to support the whole organization and its management.
    o GSS (Group Support Systems) – it facilitates the cooperation between ad hoc and permanent work groups both within an organization and between different organizations.
    o ISS (Intelligent Support System) – it is mainly designed to support the work of employees performing mental work.
    o Applications supporting production activities: CAD/CAM (Computer Aided Design/Computer Aided Manufacturing) – they are designed to support planning and production processes by using information technology devices (Shaw 1991).

The information system connected to an organization, or a part of it, provides methods to fix, process and make the information available, thereby helping the company reach its goals.

The categorization of the information systems is a difficult task as there is no unified classification. Because of the continuous development we have to classify these systems according to different point of views. We will show some clustering methods below.

According to the supported function we can distinguish:

- Office Automation Systems (OAS),
- Communication Systems,
- Transaction Processing Systems (TPS),
- Management Information Systems (MIR, MIS),
- Executive Information Systems (EIS),
- Decision Support Systems (DSS),
- Implementation Information Systems,
- Collaboration Systems.

According to the roominess of the user groups we can separate:

- Unique, special needs systems.
- Public, complex systems with general purpose.

According to the role of the user, the below types can be delimited:

- Implementation information system: creating those data, information, and documents which are necessary for doing routine tasks.
- Management information system: handling information, which is necessary for the successful and efficient decision-making activities.

III. THE METHOD OF THE RESEARCH

The statements of the paper are based on a database of a primer survey research. The survey was carried out between May and December 2012.

The questions of the questionnaire are scanning more areas by the companies, but we will focus on one major topic later. The introductory questions aim at the background information of the enterprises who filled in the questionnaire, then the IT infrastructure, internet usage habits, the practice of information management are also the subject of the inquiry. This article deals with the use of information systems and examines these questions of the questionnaire.

*A. The presumption of the research*

The purpose of our present survey is to justify our next assumption: the level of information development at Slovakian enterprises – considering every size classes – is higher than at Hungarian partners.

*B. The combination of the sample*

The target group of the survey included hundreds of micro-, small-, medium- and big enterprises. We received 94 pieces of questionnaires from Hungary and 86 pieces from Slovakia. The 21% of the Hungarian responder companies are micro-, 29% are small-, 29% are medium- and 21% are big enterprises. In spite of this more than half of the 86 Slovakian responders (51%) are micro-, 26% are small-, 15% are medium-, and only 8% are big companies. By the evaluation of the questionnaire this asymmetry in the size of the companies has not given rise to confusion, because we have compared the enterprises by countries and by size classes.

IV. COMPARISON OF THE IT DEVELOPMENT

The listed ERP systems in the questionnaire have been distinguished according to Dobay's [8] typing. The enterprises could select between three options: the system is used; not used, introduction is planned; not used, and introduction is not planned either. Because of the low rate of the second option and the easier transparency we have compared graphically only the used systems. The 1st figure illustrates the result.

The variegation of the first figure shows, that the companies use one of the listed IT systems almost in every category. However, the small and the bigger enterprises use different types by different frequency. An average micro enterprise has no need for an Executive Information System, Decision Support System or Management Information System, these do not appear on the Hungarian bar chart. The situation is different in Slovakia, the small companies also chose EIS, DSS or MIS. In the case of both countries we can say that the decision support and management systems get a higher emphasis by the increase of the company size, but this tendency is also true in every type of the systems. The Slovakian enterprises use the listed systems more often in three size classes.

The Transaction Processing Systems have been used more often by the Slovakian companies, the Hungarian data is higher only in the biggest size class. The Slovaks prefer the Office Automation Systems, while the Hungarian SME-s use the Management Information Systems more frequently than the Slovaks. The Executive Information Systems have been used more often by the Hungarian medium and large enterprises.

In case of the Decision Support Systems Hungary leads only on the medium size category. Besides the Intranet communication is more widespread among the domestic companies than in Slovakia. In total the most popular information technology systems are the Transaction Processing Systems, the Office Automation Systems and the Intranet communication, these are used by companies almost in every size class and in both countries.

Those enterprises, which use a sort of information technology systems typically, have bought ready-made systems, as you can see on the 2nd figure. This asymmetry toward the ready-made systems is the most conspicuous in Slovakia, but in Hungary only the micro size enterprises use own developed information systems mostly. The usage of the own developed systems has also appeared at the Slovakian micro companies, while the Hungarian medium and large companies use ready-made and own developed systems equally. The benefit of the latter is the customization, although the development costs are higher than in the case of the purchased systems.

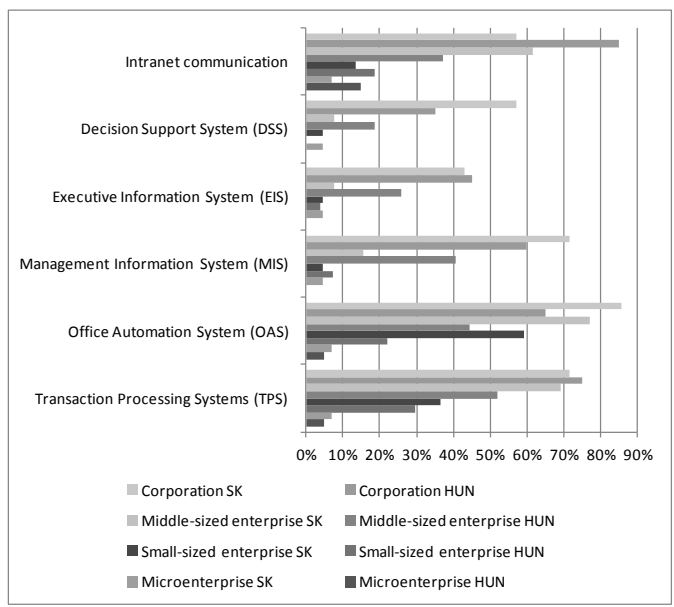

Fig. 1. Types of the applied information technology systems

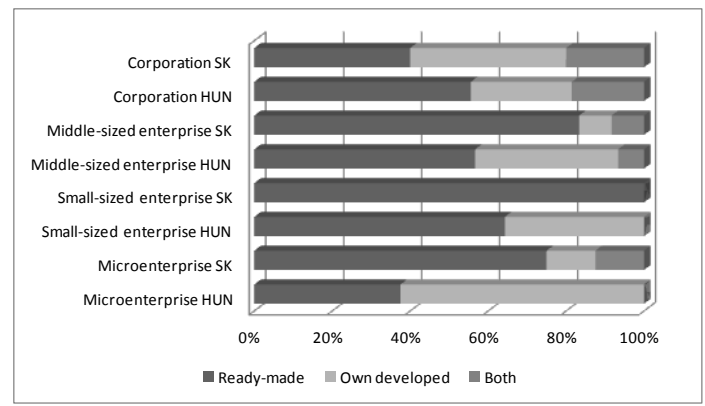

Fig. 2. Ready-made and own developed information systems

During the analysis of the information technology systems we should examine that in which area of the company operations the responders apply (at least once a week) the given software; the 3rd figure shows the result. We can see the correlation between the company size and the applied systems. The larger a company is, the more functional areas have access to the information systems. The Hungarian micro and small size enterprises use these applications in fewer areas, than the Slovaks. In contrast Hungary utilizes the software in more corporate functions in the category of medium and large companies. The most popular areas of applied information technology systems are accounting, finance and sales without reference to the company size. Logistics, marketing and senior management decision support functions also come into view by the growth of the company size.

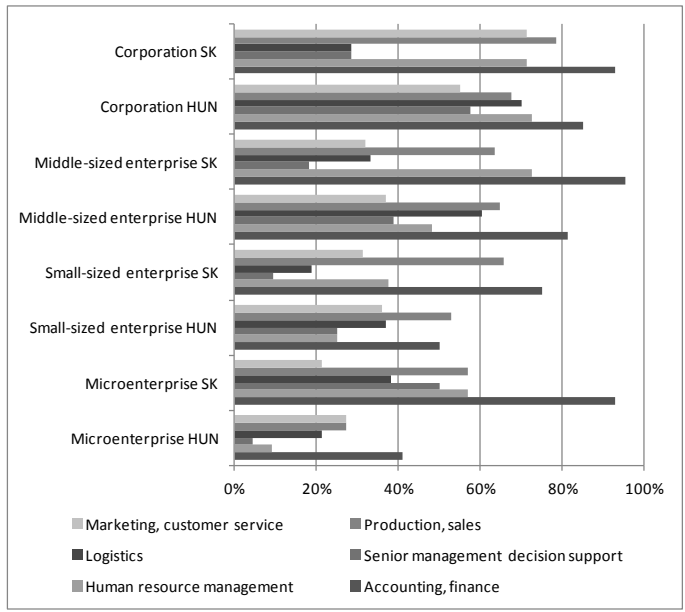

Fig. 3. Application of the information technology systems under the company operations' different areas

## V. CONCLUSION

The purpose of writing this article was to find out if Slovakian enterprises are more developed informatically than Hungarians or not. We have applied the results of an empirical survey. The subjects of the questionnaire were nearly 100 per countries.

We have waited for a verification of our assumption from the evaluation, the questions verified it partly, but some of them denied it. In case of both countries we can say that the usage of the information technology systems becomes more frequent and varied by the increase of the company size. The Slovakian companies apply ready-made systems primarily, while in Hungary the rate of the own developed and ready-made systems is almost equal. This perception refers to a higher level of IT development. The larger a company is, the more functional areas have access to the information systems. Even though micro- and small companies utilize the IT softwares in more functional areas in Slovakia, in the category of medium- and large companies the Hungarian responders use these applications in a wider range of activities [20].

To sum up, according to the survey we can say that there is no significant difference between the IT development of Hungary and Slovakia. At the same time it seems that there is a directly proportional relationship between the company size and the willingness to use the information technology systems.